\documentclass[sigconf,nonacm=true]{acmart}

\usepackage{caption}
\usepackage{enumitem}
\usepackage{subcaption}
\usepackage{balance}
\usepackage[ruled,vlined]{algorithm2e}

\AtBeginDocument{%
  \providecommand\BibTeX{{%
    \normalfont B\kern-0.5em{\scshape i\kern-0.25em b}\kern-0.8em\TeX}}}

\begin{document}

\title{Optimal Resource Allocation for Serverless Queries}

\author{Anish Pimpley}
\authornote{Both authors contributed equally to this research.}
\affiliation{%
  \institution{Microsoft}
  \country{}
}

\author{Shuo Li}
\authornotemark[1]
\affiliation{%
  \institution{Microsoft}
  \country{}
}

\author{Anubha Srivastava}
\authornote{Work done while at Microsoft.}
\affiliation{%
  \institution{Two Sigma}
  \country{}
}

\author{Vishal Rohra}
\affiliation{%
  \institution{Microsoft}
  \country{}
}

\author{Yi Zhu}
\affiliation{%
  \institution{Microsoft}
  \country{}
}

\author{Soundararajan Srinivasan}
\affiliation{%
  \institution{Microsoft}
  \country{}
}

\author{Alekh Jindal}
\affiliation{%
  \institution{Microsoft}
  \country{}
}

\author{Hiren Patel}
\affiliation{%
  \institution{Microsoft}
  \country{}
}

\author{Shi Qiao}
\affiliation{%
  \institution{Microsoft}
  \country{}
}

\author{Rathijit Sen}
\affiliation{%
  \institution{Microsoft}
  \country{}
}

\renewcommand{\shortauthors}{Pimpley and Li, et al.}

\begin{abstract}
Optimizing resource allocation for analytical workloads is vital for reducing costs of cloud-data services. 
At the same time, it is incredibly hard for users to allocate resources per query in serverless processing systems, and they frequently misallocate by orders of magnitude.
Unfortunately, prior work focused on predicting peak allocation while ignoring aggressive trade-offs between resource allocation and run-time. Additionally, these methods fail to predict allocation for queries that have not been observed in the past. In this paper, we tackle both these problems. We introduce a system for optimal resource allocation that can predict performance with aggressive trade-offs, for both new and past observed queries. We introduce the notion of a performance characteristic curve (PCC) as a parameterized representation that can compactly capture the relationship between resources and performance. To tackle training data sparsity, we introduce a novel data augmentation technique to efficiently synthesize the entire PCC using a single run of the query. Lastly, we demonstrate the advantages of a constrained loss function coupled with GNNs, over traditional ML methods, for capturing the domain specific behavior through an extensive experimental evaluation over SCOPE big data workloads at Microsoft.
\end{abstract}



\maketitle

\section{Introduction}
\label{sec:intro}

\begin{figure}[t]
\centering
\includegraphics[width=0.475\textwidth]{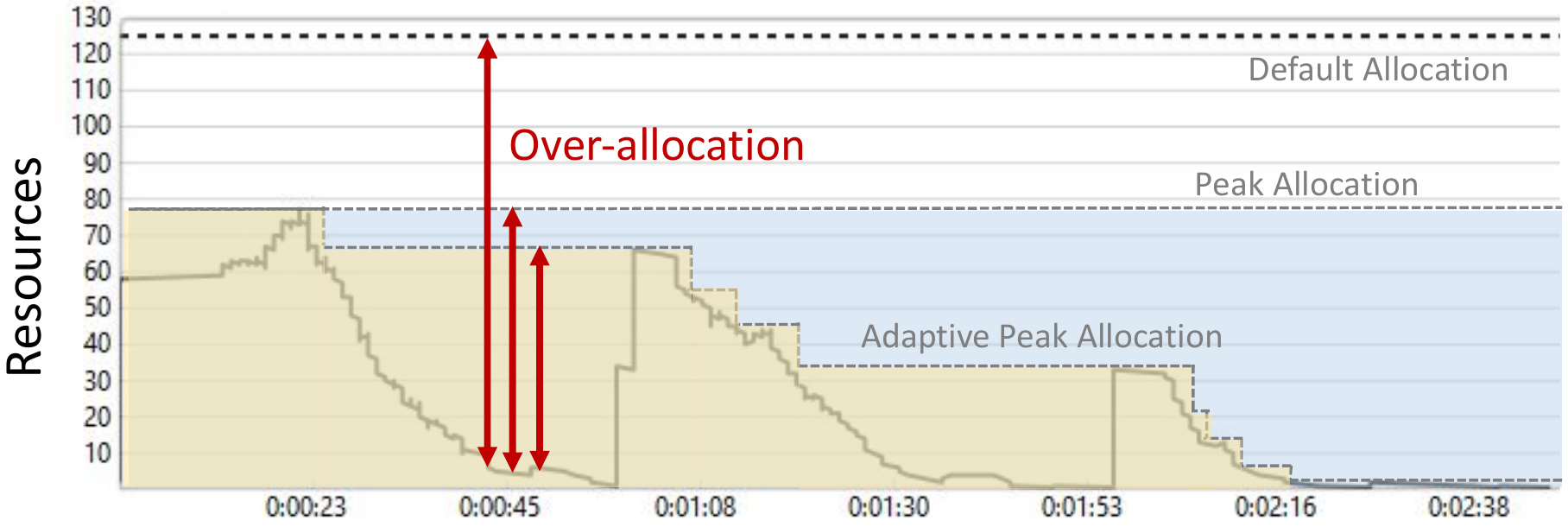}
\caption{Resource usage skyline of a SCOPE job and the over-allocation incurred by different allocation policies.}
\vspace{-0.3cm}
\label{fig:resourceSkylines}
\end{figure}

Serverless computing alleviates the need for dedicated resources in the cloud~\cite{serverless}.
For query processing, this means that 
serverless query processing platforms, such as SCOPE~\cite{chaiken2008scope}, Athena~\cite{athena}, and Big Query~\cite{tigani2014google}, can dynamically allocate resources for each incoming queries.
However, allocating resources efficiently is a challenge.
This is because analytical queries have a complex relationship between resource allocation and performance, and that relationship changes over the course of query execution~\cite{jyothi2016morpheus}.
To illustrate, Figure~\ref{fig:resourceSkylines} shows the resource consumption skyline of a SCOPE query (called {\it job}) on the Cosmos big data analytics platform at Microsoft~\cite{chaiken2008scope}. 
As evident from Figure~\ref{fig:resourceSkylines}, it is incredibly hard for a user to anticipate such a complex resource consumption behavior and allocate resources accordingly. Recent works have considered either predicting the peak resource consumption~\cite{sen2020autotoken} or adaptively giving up non-required resources over the course of query execution~\cite{tokenshaper}.
However, we also see numerous valleys in the resource consumption skyline in Figure~\ref{fig:resourceSkylines}. Therefore, allocating for the peak resource consumption, either upfront or adaptively over time, is highly conservative and misses valuable opportunities for operational efficiency. 

\begin{figure}[t]
\centering
~\includegraphics[height=1.5in]{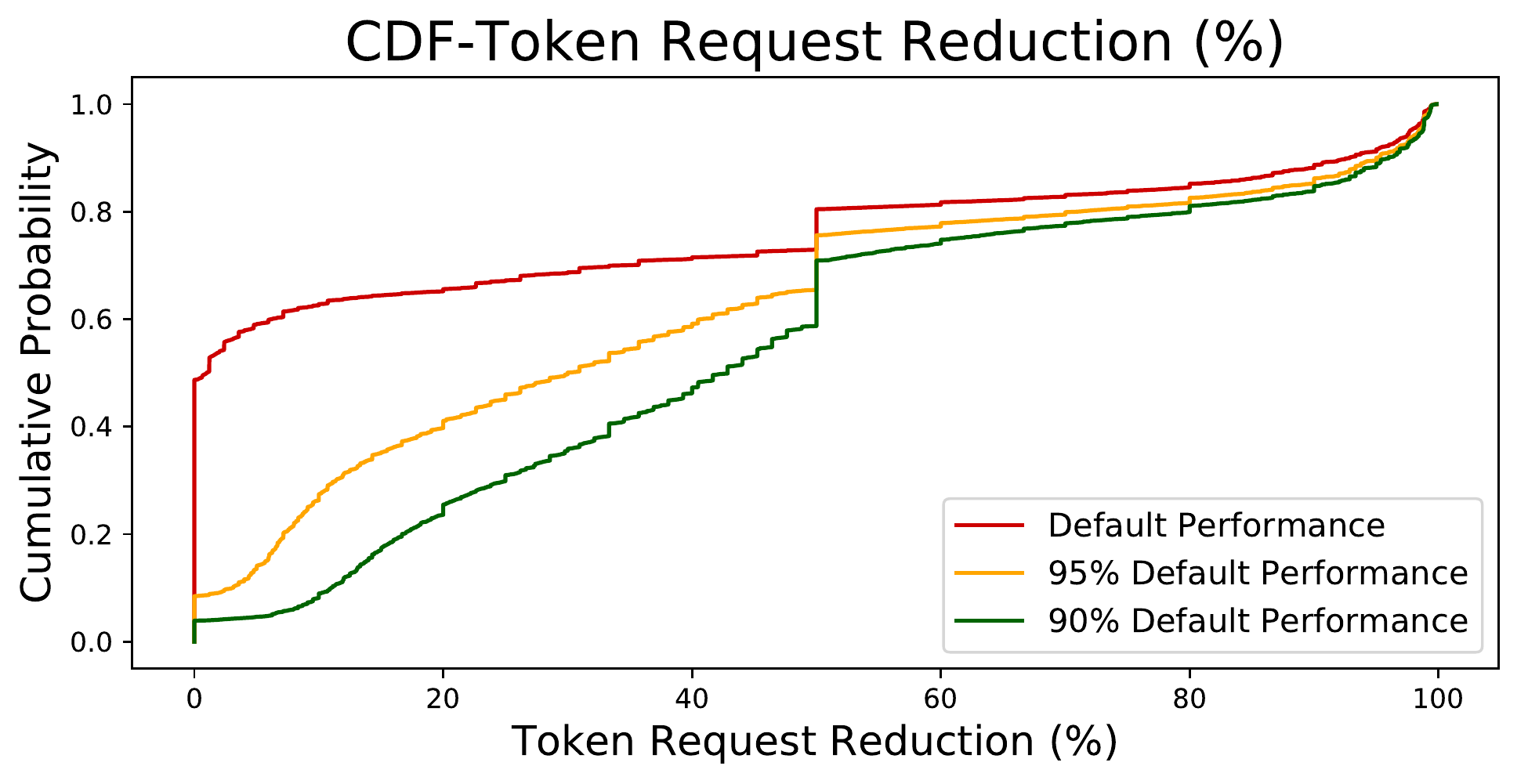} 
\vspace{-0.3cm}
\caption{Cumulative distribution function of potential token request reduction in SCOPE jobs.}
\vspace{-0.3cm}
\label{fig:cdfofTokenSavings}
\end{figure}

To illustrate the opportunity with a more aggressive resource allocation strategy,
Figure~\ref{fig:cdfofTokenSavings} shows the cumulative distribution of potential reduction in containers (also referred to as {\it tokens}) requested by SCOPE jobs.
We can see that more than $50\%$ of the SCOPE jobs could request fewer tokens without any (estimated) impact on performance, with $33\%$ of the jobs potentially needing less than $25\%$, and $20\%$ needing less than $50\%$ of the tokens (red curve).
Additionally, if the users are willing to accept a $5\%$ performance loss, then the percentage of jobs that could utilize fewer tokens go up to $92\%$, with $30\%$ needing less than $50\%$ tokens (yellow curve).
Utilizing fewer tokens reduces job wait time and improves the overall resource availability for other jobs in the cluster~\cite{sen2020autotoken}.
Interestingly, our conversations with internal Microsoft users reveal that they have the above intuition and are looking to allocate resources below the peak. However, it is very hard for them to accurately estimate the optimal resources for different jobs~\cite{10.1145/2987550.2987566}. To understand further, we ran a qualitative user study with 12 engineers. Only 5 users demonstrated an understanding of tokens and token allocation, and all 5 of them either guessed or selected the default when deciding the number of tokens to allocate per SCOPE job.

\begin{figure}[t]
\centering
~\includegraphics[height=1.5in]{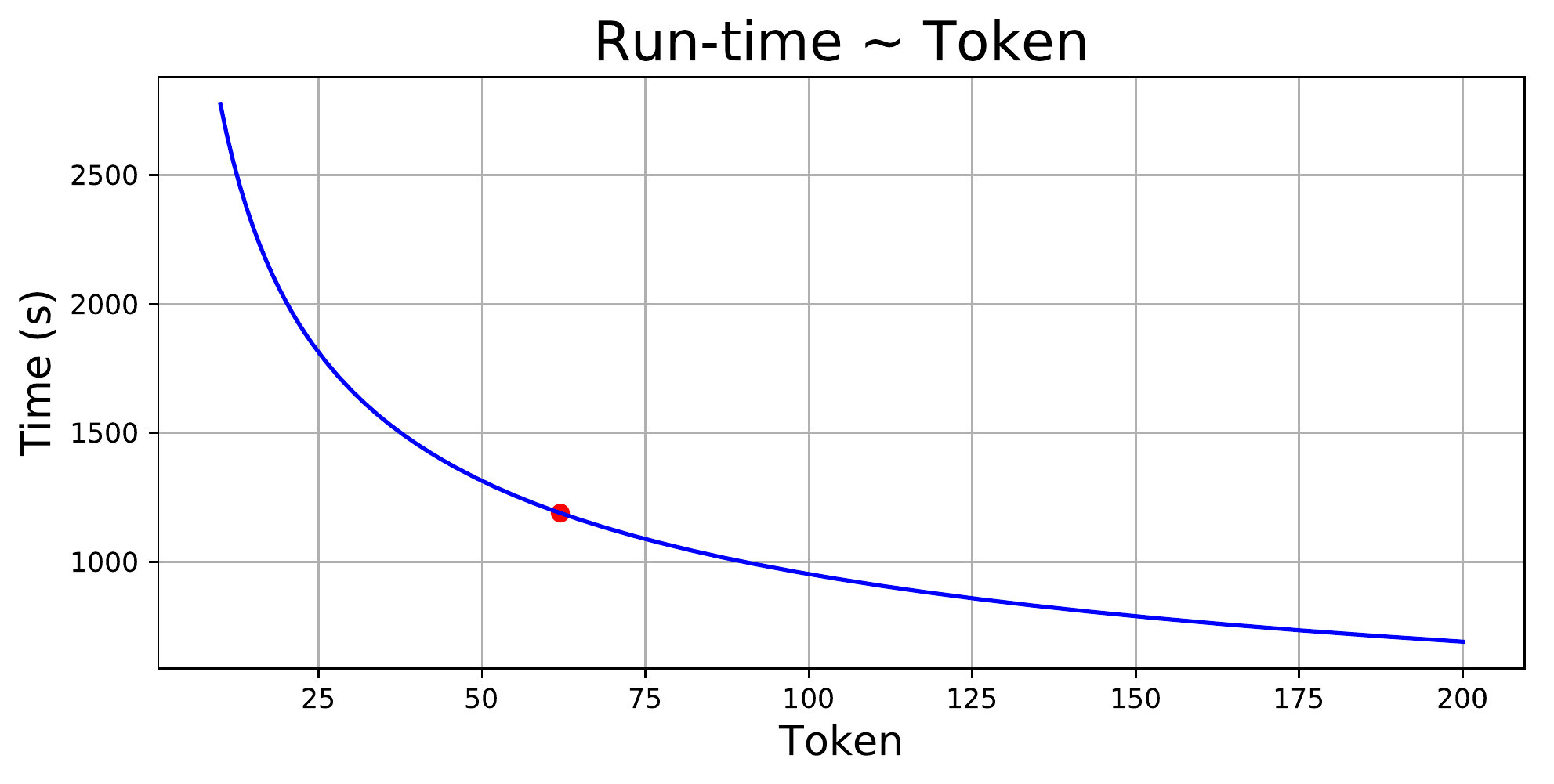}
\vspace{-0.3cm}
\caption{Trade-off between allocated resources and job run-time. Red marker indicates an elbow in the curve.}
\vspace{-0.3cm}
\label{fig:tokenRuntimeTradeoff}
\end{figure}

In this paper, we present a machine-learning based approach to select, at compile-time, the optimal resources for each query in a serverless query processing system. Our key observation is that the relationship between resources and performance could be approximated using an exponentially decaying curve, shown in Figure~\ref{fig:tokenRuntimeTradeoff} and referred to as {\it performance characteristic curve} (PCC) henceforth.
This is because while job performance improves significantly with more tokens initially, the change is diminishing for larger token counts. Our goal then is to learn the PCC for all jobs, i.e., given the compile time characteristics or features of a job, predict the parameters for its PCC.
Such a curve can be used for both point prediction, i.e., predict the job run-time given a token amount, and trend prediction, i.e., predict the job run-times for each point in a range of token counts. 
The monotonicity property of PCC is further useful in helping users understand the proposed model and the performance/resource trade-offs, allowing them to tune the resource allocation based on their acceptable performance range and service-level objectives (SLOs). 

\noindent{\bf Challenges.}
The biggest challenge in learning the PCC of all jobs is the
limited historical data available for training. Queries in the historical workloads are executed with a single token count, while we need to track performance changes across different token counts to train the PCC. The problem becomes worse with high diversity and massive scale in typical enterprise workloads that are hard to model, e.g., SCOPE processes hundreds of thousands of jobs per-day over petabytes of data from several different business units across the whole of Microsoft.
One option could be to use a job's most recent resource allocation skyline to estimate PCC, however, the skyline could change significantly over time due to changes in workloads, such as input sizes. Furthermore, new or ad-hoc jobs with no historical data do not have historical skylines. 
Finally, unlike prior approaches for learning cost models~\cite{costLearner}, the PCC model needs to capture the diminishing nature of performance improvements as more resources are allocated.

\noindent{\bf Contributions.} We address the above challenges in this paper and make the following key contributions:
\vspace{-0.1cm}
\begin{itemize}[leftmargin=10pt]
\item 
  We present TASQ (\underline{\bf T}oken \underline{\bf A}llocation for \underline{\bf S}erverless \underline{\bf Q}ueries), an end-to-end ML pipeline to predict optimal token counts in SCOPE like serverless query processing systems. (Section~\ref{sec:system-overview})
\item We introduce an efficient data augmentation technique for enriching sparse training data using a job skyline simulator, coined AREPAS, that can accurately synthesize skylines with alternate token counts. (Section~\ref{sec:dataaug})

\item We describe an ML approach for point and trend predictions using a rich set of features from past job graphs. Our model characterizes PCC using two parameters of a power-law curve and predicts these parameters for a given job. (Section~\ref{sec:performance-models})
  
\item We present the results of extensive experimental evaluation of TASQ incorporating different ML approaches.
Our results show that an XGBoost-based approach has the best accuracy for point prediction. However, its trend prediction is not always correct. In contrast, feed forward neural networks and graph neural networks guarantee a monotonically non-increasing trend between resources allocation and job run-time, and both models have reasonably good accuracy for point prediction~(Section~\ref{sec:evaluation}). 
\end{itemize}

\section{TASQ Overview}
\label{sec:system-overview}

We study the problem of efficient resource allocation in the context of SCOPE~\cite{chaiken2008scope}, which runs the internal big data workloads at Microsoft and comprises of hundreds of thousands of jobs per-day, each being a DAG of operators and executed using thousands of containers (referred to as tokens) in parallel, and processing petabytes of data.
SCOPE users submit their analytics jobs along with a token amount that is allocated as guaranteed resources before the job can start. Unfortunately, users rarely make informed decisions on the appropriate number of tokens needed for their jobs, and they typically opt for the default values.
Poor token allocation leads to performance loss, higher wait times, and more operational costs for customers. 
Our recent work considers peak resources, either upfront or adaptively over time~\cite{sen2020autotoken,tokenshaper}. However, peak allocation is utilized only for a small portion of a job's lifetime. Therefore, we could allocate below peak without incurring a noticeable loss in performance. We refer to this maximal allocation under the peak, that optimally trades off resource cost and run-time as the optimal resource allocation.

\subsection{Problem Definition} 

Let $J$ be an analytical job and $A$ be the resources (token count) allocated to it. Also, let $R$ be the performance, measured typically in terms of the total run-time of job $J$. Our goal then is to learn the performance characteristic curve of job $J$, i.e., its run-time as a function of resource allocation, as illustrated in Figure~\ref{fig:tokenRuntimeTradeoff}. Formally, 
\vspace{-0.1cm}
\begin{equation} \label{eq:pcc}
    PCC_J: R = f(A)
\end{equation}
Given a $PCC_J$, we can then find the optimal resource allocation using gradient descent with a termination condition, i.e., a threshold beyond which the gains in performance are diminishing. This threshold could be controlled by users, e.g., minimum $1\%$ performance improvement for every additional token count.

\subsection{System Implementation} We have implemented TASQ for optimal resource allocation, within the broader workload optimization platform that we have been building at Microsoft~\cite{peregrine}.
The TASQ pipeline starts from the workload repository that includes query plans, run-time characteristics, and other job metadata, 
and transforms them into a clean model-suitable format, which is then stored on Azure Data Lake Storage (ADLS)~\cite{adls}. Thereafter, TASQ runs the featurization, training, and inferencing steps on Azure Machine Learning (AML)~\cite{aml}. The final model is then registered in AML model store and deployed as an Azure Kubernetes Service (AKS)~\cite{aks}. This service endpoint is used to integrate with a Python client for SCOPE. For an incoming SCOPE job, the client fetches the compile time features, transforms them into inference-suitable format, and passes them to the deployed model service.  The model service returns the parameters of the PCC that are required to make the token prediction. 
Based on the configuration, the client can either directly use the predicted optimal token count for execution or display the PCC for users to understand the tradeoff and make an informed decision.

TASQ uses a novel simulator, called AREPAS, to generate augmented training input.
Given a job's resource consumption skyline, AREPAS can synthesize run-time for the same job with different token allocation. We utilize AREPAS to generate synthetic data that augments the historical dataset. 
To train the model, 
we characterize the output of the simulator by distilling it down to two parameters that define a power-law shaped PCC curve. 
Our training setup uses these parameters as targets for the ML model. This implicitly constraints the model and guides it to learn relationships that reflect known intuitions of domain experts.  
To accurately learn these parameters, we construct a 
loss function containing two components which are balanced by tuned weights, namely mean absolute error of PCC parameters and mean absolute error of run-time prediction.

In the rest of the paper, we first describe training data augmentation in Section~\ref{sec:dataaug}, then we discuss the prediction models in Section~\ref{sec:performance-models}, and finally show the experimental results in Section~\ref{sec:evaluation}. 

\section{Data Augmentation}
\label{sec:dataaug}

Our goal is to model the relationship between job run-time and
allocated resources (tokens), given the compile-time job characteristics. 
However, to train such a model, we need
the run-times for the same job with different token counts. Unfortunately,
historical data only contains the job run-time at
one single token count (the one actually used by the job). 
Therefore, below we describe our approach to augment the training dataset.

\subsection{Challenges}

Augmenting the job telemetry that we use for training is challenging because each of the past jobs ran with a single token count, and their performance over other token counts is unobserved and unknown to us. Unfortunately, traditional data augmentation approaches such as GANs~\cite{goodfellow2014generative} will not work in our scenario. This is because GAN models generate new samples based on 
what we already observe from historical data, while we want to produce telemetry at other token counts for each job. 
Furthermore, building GAN models to generate different resource consumption skylines
is as complex as modeling run-time as a function of token allocation, albeit with limited training data. 
Instead, we want to 
keep data augmentation a lightweight process that can also work for newer jobs.

\subsection{AREPAS}

We introduce \emph{\textbf{Are}a
\textbf{P}reserving \textbf{A}llocation \textbf{S}imulator} (\textbf{AREPAS}) to augment the training data in TASQ using system-level intuition. 
Specifically, AREPAS assumes the total amount of work or the area under the resource consumption skyline remains fixed.
We discretize the resource consumption skyline at 1 second's granularity, a 1x1 square in the plot represents 1 token-second, and the total number of squares under the resource consumption skyline is the total amount of token-seconds used by the job.
We then assume that the total token-seconds stay constant.
Note that AREPAS simulates performance at the coarse-grained granularity of a job and its resource consumption skyline, instead of modeling it for every stage level in the job execution graph as in Jockey~\cite{ferguson2012jockey}.
This is because a stage-level simulator needs to make assumptions about the scheduler that is much more unpredictable. Instead, AREPAS simply assumes that the total amount of work remains constant.

\begin{figure}[t]
\centering
    \begin{subfigure}[b]{0.235\textwidth}
         \centering
         ~\includegraphics[width=\textwidth]{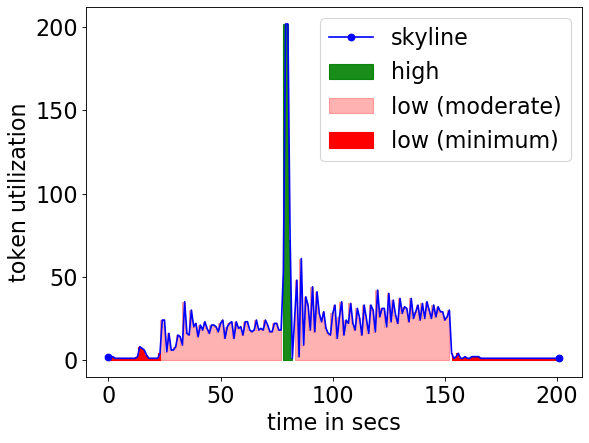}
         \caption{Peaky Skyline}
         \label{fig:peaky}
    \end{subfigure}
    \hspace{-0.2cm}
    \begin{subfigure}[b]{0.235\textwidth}
         \centering
         ~\includegraphics[width=\textwidth]{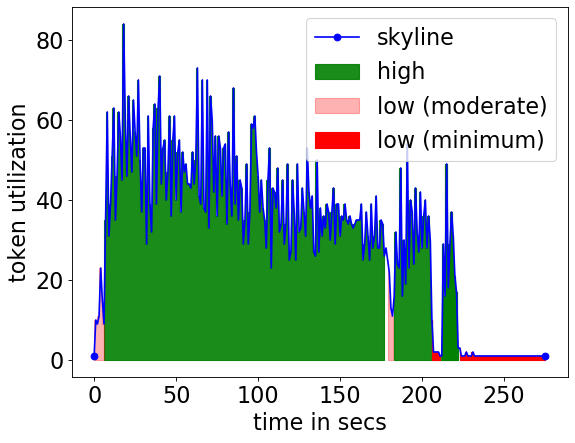}
         \caption{Flatter Skyline}
         \label{fig:flatter}
     \end{subfigure}
\vspace{-0.2cm}
\caption{Resource skylines divided into sections by resource consumption.}
\vspace{-0.4cm}
\label{fig:peakyflatskyline3sections}
\end{figure}

\begin{figure}[t]
\centering
~\includegraphics[width=0.24\textwidth]{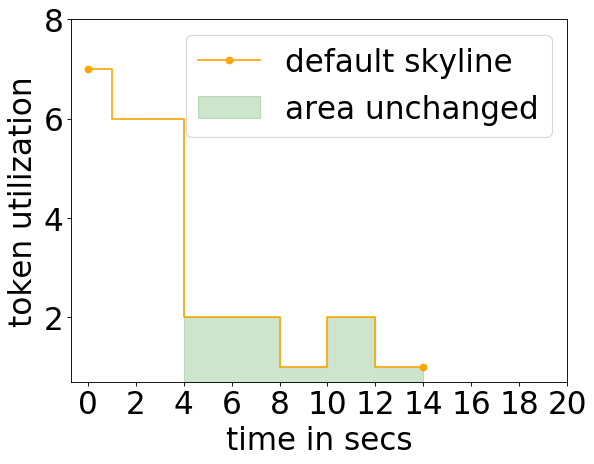}
~\includegraphics[width=0.23\textwidth]{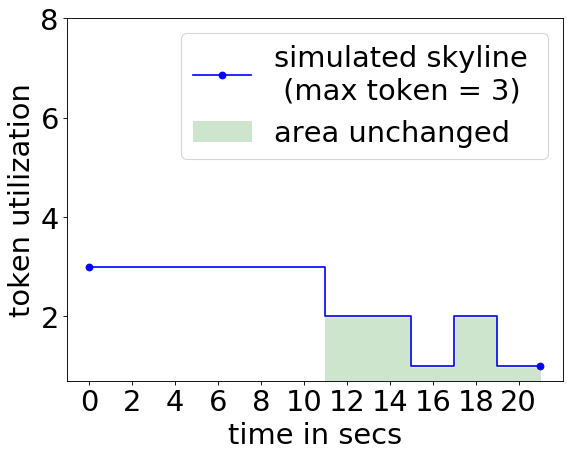}
\vspace{-0.5cm}
\caption{Unchanged section between the Ground truth (left) and simulated (right) skylines.}
\vspace{-0.5cm}
\label{fig:unchangedsection}
\end{figure}

\begin{figure}[t]
\centering
~\includegraphics[width=0.24\textwidth]{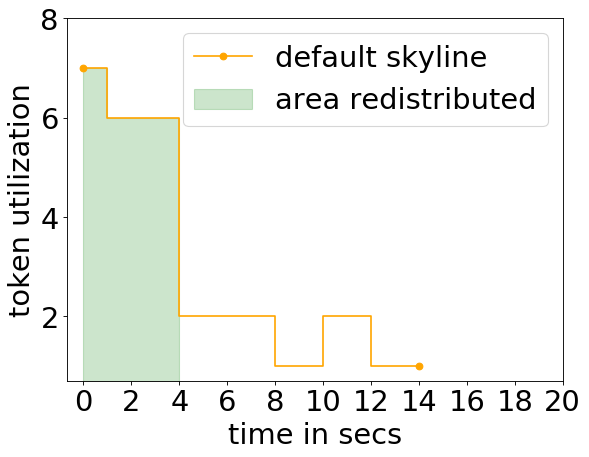}
~\includegraphics[width=0.24\textwidth]{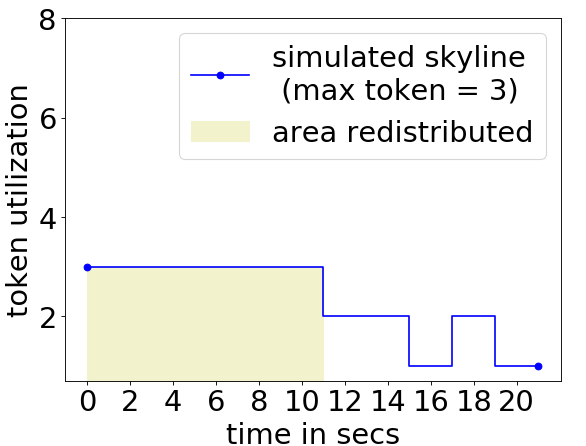}
\textit{The simulator reallocates work with fewer resources.
To preserve area, the reallocated portion in the simulated skyline takes more than twice as long, when allocated a little less than half as many tokens.}
\vspace{-0.1cm}
\caption{Redistributed section between the Ground truth (left) and simulated (right) skyline.}
\vspace{-0.5cm}
\label{fig:changedsection}
\end{figure}

Figure~\ref{fig:peakyflatskyline3sections} shows two examples of different types of skylines. 
In both cases, we observe that a significant
portion of the run-time is spent utilizing a fraction 
or at times the near minimum (shown as red regions)
number of tokens. The peakier job (Figure~\ref{fig:peaky}) spends longer in the pink/red regions and
flatter job (Figure~\ref{fig:flatter}) spends longer time in the green region.
We want to simulate the skyline for the above jobs at any token
allocation value which is lower than the original token allocation.
Additionally, we aim to achieve this using only the above skyline as an input to the simulator.
To achieve this, we make the following assumptions:

\begin{itemize}[leftmargin=10pt]
\item
  The simulated skylines are deterministic, i.e., for any given combination of a job and a token allocation we always get the same skyline. Thus, we do not model any stochastic behavior in the execution environment, e.g., random failures of machines, cluster load at the time of execution, noisy neighbors on same physical machines, scheduler queues, etc.
\item
  Each 1x1 block under the skyline, representing 1 token-seconds each, is independent of each other. So, if a compute process consumes 10 token-seconds over its lifetime, then it will take 1 second and 10 seconds to complete with 10 tokens and 1 token respectively, based on the right-nearest integer approximation. 
\item
  Sections of a simulated skyline that are below the
  allocated resources will have the shape of their skylines unchanged when simulated. (Figure~\ref{fig:unchangedsection}) 
\item
  For sections of a skyline that go over the allocated resources, the simulator adapts the skyline's shape while keeping the total amount of work done in those sections constant. (Figure~\ref{fig:changedsection})
\end{itemize}

Using the above assumptions, we simulate skylines over different resource allocation for the same job. For simulation, the skyline is divided into sections, where each section is a contiguous chunk of the skyline that is completely under or completely over the new allocation.
Figure~\ref{fig:unchangedsection} shows how over-allocated sections (where usage $<$ new allocation) are copied without change to the new skyline.
Figure~\ref{fig:changedsection} shows how 
Sections of the skyline that are under-allocated (cut-off by new allocation) are immediately added back as a task in
front of the part that was cut off. This pushes the rest of the plot
forward and as a result increases run-time. The area that is being added is the same as the area that is being cut
out, as per the area preservation design choice.

\begin{algorithm}[h]
\SetKwFor{RepeatTimes}{repeat}{times}{end}
\SetKwInOut{Input}{input}\SetKwInOut{Output}{output}
\Output{$Ssim$: simulated skyline $(list)$ ;}
\Input{$Sog$: original skyline $=[s_{0},s_{1},s_{2},.....s_{runtime}]$\\
 $Nt$: new allocation threshold to simulate $(int)$\; }
 
$sectionStartIDs = []$\\
\For{$i\leftarrow$  $1$  \KwTo  $runtime$ }
{
\lIf{$sign(Sog[i]-$Nt$) \neq sign(Sog[i-1]-$Nt$)$}
{$sectionStartIDs.append(i)$}
}
$sections = Sog.split(sectionStartIDs)$:

\For{sec in sections}
{
 \eIf{$sec[0]$ > $Nt$}
 {
 $secArea = sum(sec)$\\
 $newSecLength = int(secArea/Mt)$\\ 
 \RepeatTimes{$newSecLength$}{ $Ssim.append(Mt)$}
 }
 {
 $Ssim.appendAllElements(sec)$
 }
}
 \Return{$Ssim$}
\caption{How to simulate a skyline}
\label{alg:algorithm}
\end{algorithm}

Algorithm \ref{alg:algorithm} follows a few simple steps. First, we identify the timestamps where the skyline intersects with the new allocation. Then, we use the time stamps to divide the skyline into sections. Sections that are above the new allocation are lengthened until they can fit under the new allocation while keeping their area constant. Sections that are under the new allocation stay unchanged and are copied over as is. These new sections are then stacked in front of the other to obtain the newly simulated skyline. 

\begin{figure}[t]
\centering
~\includegraphics[width=0.24\textwidth]{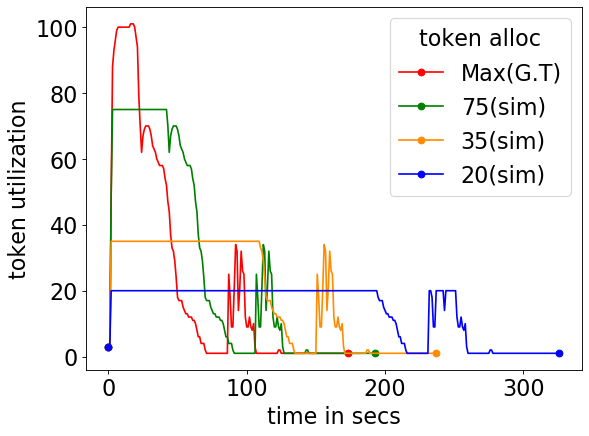}
~\includegraphics[width=0.24\textwidth]{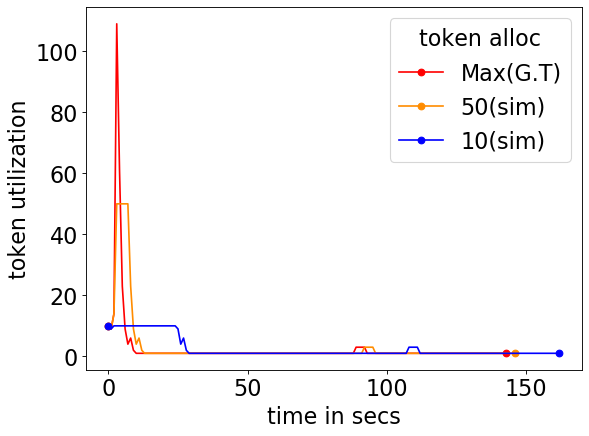}\\
\vspace{-0.2cm}
\caption{Flatter jobs (left) lose significant performance as soon as the token allocation is decreased, while peaky jobs (right) can tolerate a significant loss of resource allocation before they lose performance.}
\vspace{-0.4cm}
\label{fig:peakyflatlosstolerance}
\end{figure}

Figure~\ref{fig:peakyflatlosstolerance} shows the simulation of two kinds of jobs (peaky and flatter) for different token allocation, with the simulated skylines shown in different colors. Note that the peaky jobs show less impact on performance with lower resource allocation than more flatter jobs. This is expected since peaky jobs have deeper valleys and hence more aggressive allocation can shift some of the work to other low activity regions, thus freeing up resources for other jobs.
As shown later in experiments (Section~\ref{subsec:simulator-validation}), the run-time produced from simulated skylines are only up to $50\%$ off in the worst case. 

\section{Prediction models}
\label{sec:performance-models}
In this section, we first define the monotonicity constraint of trend predictions. Then we explain why we prefer a global model over fine-grained models, how we featurize the job metadata and use ML models to predict job run-time as a function of token allocation. 

\subsection{Monotonicity constraint}
\label{subsec:monotonicity-constraint}

Our goal is to fit a function to the PCC, i.e., the relationship between job run-time and token allocation. From Amdahl's law, the parallelizable portion of a workload has an
inverse relationship with run-time~\cite{amdahl1967validity}. However, the steepness of the inverse
relationship is not captured by a simple inverse proportionality. To
generalize this idea, we characterize the inverse relationship as power-law:  
\begin{equation}
Runtime = f\left(A:Token Allocation\right) = b \times \text{A}^{a}
\end{equation}

\begin{figure}[t]
\centering
~\includegraphics[width=0.24\textwidth]{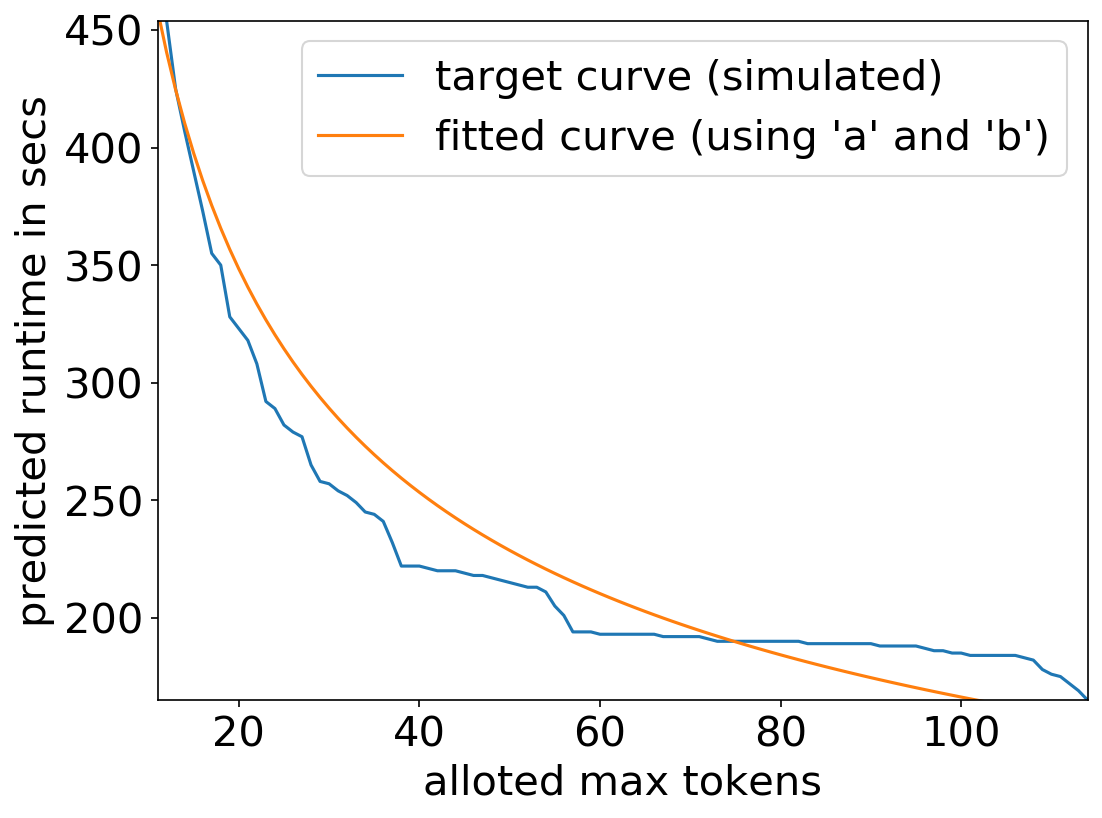}
~\includegraphics[width=0.24\textwidth]{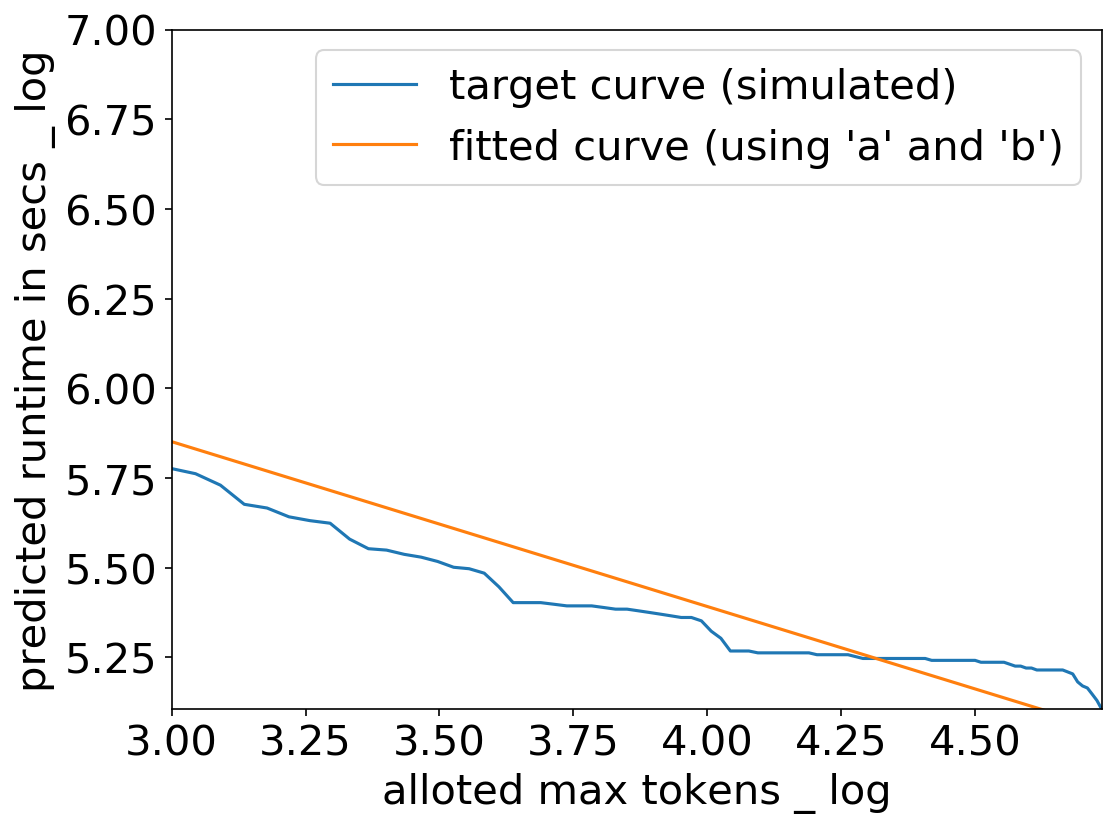}
\vspace{-0.4cm}
\caption{Performance characteristic and the fitted curve in absolute and log-log spaces}
\vspace{-0.4cm}
\label{fig:perfcharacurve}
\end{figure}

Where `a' and `b' are the 2 scalar parameters of the curve. Amdahl's law
can be seen as a special case where `a' = -1. Thus, the run-time versus token relationship would be monotonically increasing if the signs of `a' and `b' are consistent and decreasing if the signs of `a' and `b' are inconsistent. 
Since power-law relationships can be represented as a linear curve in
the log-log space, we transform the above equation accordingly and fit a
straight line through it. Figure~\ref{fig:perfcharacurve} visualizes the simulated curve and fitted power-law curve in
both spaces. 

The above approach has several advantages.
First of all, using just the two parameters for fitting PCC
helps represent the domain knowledge compactly, which not only simplifies the problem but also makes it easier for users to understand the trade-offs given a PCC. Learning a model that can predict the two parameters with good accuracy, also guarantees the correct shape of PCC. In contrast, predicting the run-time at a bunch of token amount and joining the predictions will not necessarily reach a monotonically non-increasing PCC (as we show in section~\ref{subsec:model-validation-full}).
Furthermore, we can generalize
the relationship between job metadata at compile time and performance
characteristic to any job, including previously unseen jobs. 

A few pitfalls can be expected when using simulated data directly as training input. The simulated data is entirely a function of the originally observed points, which violates the assumption about training samples being independent. This makes the simulated data a second-class citizen in the training input and risks teaching the model how to better estimate the AREPAS simulator instead of the run-time performance itself. Fortunately, using our modeling approach allows us to mitigate this to some degree. This is because, one component of the loss discussed in Section~\ref{subsec:loss-function} only uses ground truth for learning. Thus, the simulator serves as a tool for constraining the model's architecture using inductive bias instead of explicitly guiding it.

We applied and compared three models: XGBoost, feed forward neural networks (NN) and graph neural networks (GNN) in this paper. All these models can be used to make predictions for the power-law curve parameters (Section \ref{subsec:model-training}), that leads to either monotonically increasing or decreasing PCC. However, we are able to enforce monotonically non-increasing curve by design for NN and GNN (Section \ref{subsec:loss-function}). Below we first discuss the granularity of our models and the featurization, before discussing the models.

\subsection{Generalized global model}
\label{subsec:global model}
We consider two learning granularities: (1)~global model: a single model for all incoming jobs, both recurring and ad-hoc jobs, and (2)~fine-grained models: grouping similar jobs together and training separate models for each group, similar as in prior work~\cite{sen2020autotoken}. Compared to global model, the fine-grained approach could improve the accuracy by specializing each subgroup's run-time versus token relationship. 
However, a global model can provide coverage for all jobs, whereas the coverage for the fine-grained approach is limited to recurring jobs seen in the past. Given that token allocation is difficult for the users, we want to predict resource allocation for all incoming jobs and not be restricted to recurring workloads. Therefore, we choose the global model approach in this work,
 and as we shall show in Section~\ref{subsec:model-validation-full}, the accuracy is quite good as well. 
 
\subsection{Featurization}
\label{subsec:featurization}

\begin{table}[tb]
\caption{Featurization methods and target variables.}
\vspace{-0.3cm}
\label{tab:feat_summary}
\small
  \begin{tabular}{cll}
    \toprule
    Model & Features & Target Variables\\
    \midrule
    \textbf{XGBoost} & Aggregated Job Level & Run-time\\
    \textbf{NN} & Aggregated Job Level & PCC Parameters\\
    \textbf{GNN} & Operator Level & PCC Parameters\\
     & Graph Representation &  \\
  \bottomrule
\end{tabular}
\vspace{-0.3cm}
\end{table}

\begin{table}[tb]
\caption{Operator level features from the workload.}
\vspace{-0.3cm}
\label{tab:feat_type}
\small
  \begin{tabular}{cll}
    \toprule
    Type & Features\\
    \midrule
Continuous & Cardinality (estimated, input, input children), \\
 & Average Row Length,\\
 & Cost (estimated, estimated exclusive, estimated total)\\
Count & Number of Partition, Number of Partitioning Column, \\
 & Number of Sort Column \\
Categorical & Physical Operator (35 types), Partition (4 types)\\
  \bottomrule
\end{tabular}
\vspace{-0.3cm}
\end{table}

We now describe the featurization and modeling set-up that differs across XGBoost, NN and GNN.
Table~\ref{tab:feat_summary} shows three kinds of featurization of the query workload (consisting of query plans and their corresponding run-time metrics):

\textbf{Aggregated job-level features.} XGBoost and NN require same input features dimension for each job. As a result, we aggregate the metrics from each operator in the query plan to form an equal length feature vector for each job and feed these models with the aggregated job level features. The continuous and count variables are aggregated by mean, and the categorical variables are aggregated by frequency count. The number
of operators and stages are included as features as well. For each job,
we have a $1$ $\times$ $P_{J}$ feature vector, where $P_{J}$ is the number of aggregated features. 

\textbf{Operator-level features.} 
Unlike XGBoost and NN, GNN is flexible in that the model can take operator level features, as shown in Table \ref{tab:feat_type}, directly as input and the input feature dimension could vary by job, depending on how many operators a job has. For each job, the operator-level features are represented as a $N $ $\times$ $P_{O}$ feature matrix, where N is the number of operators and $P_{O}$ is the number of operator-level features. 
Operator-level features avoid the information loss due to aggregation. 

\textbf{Graph representation.} We represent the query plan 
graph structure by an adjacency matrix, and it is obtained from the
DAG of the operators. We use this matrix to measure the
connectivity of the nodes for the GNN.
\subsection{Model training}
\label{subsec:model-training}
We now describe training three kinds of models, namely, XGBoost, NN and GNN, to learn the parameters of PCC.

\textbf{XGBoost}~\cite{Chen_2016}: 
Since we cannot use XGBoost to predict the two PCC parameters jointly, we first train XGBoost with Gamma regression trees to predict job run-time, then form the PCC with a series of run-time predictions. To enable the model to learn a monotonically non-increasing PCC, we perform data augmentation and explore alternate points on both side of the default allocation. For each job, we generate two more observations using the AREPAS simulator at 80\% and 60\% of observed token amount. For the
over-allocated jobs, the peak token allocation is observed, and two
other observations are generated, that is 120\% and 140\% of the peak token
with run-time floored at peak allocation. With run-time prediction, two approaches are considered to form a PCC: (1) XGBoost Smoothing Spline (XGBoost SS): The predictions at multiple token amount (+-40\% of the observed token value) are smoothed to form a PCC, and (2) XGBoost Power-Law (XGBoost PL): The predictions at multiple token amount are used to fit a power-law shaped PCC. 

\textbf{Feed-Forward Fully Connected Neural Networks}:  
The job level aggregated features are fed into a multi-layer fully connected
NN to predict the two power-law shaped PCC
parameters. 
The run-time at specific token amounts can be predicted by plugging the
token values into that function. In addition, we can define the peak
token amount or the optimal token amount (e.g., at a specific token
amount, adding/reducing one token will cause the run-time to
decrease/increase by p\%). This token amount can be calculated using the
slope and run-time predictions based on the predicted PCC function as follows:
$\frac{f'(A)}{f(A)} = p\%$

\begin{figure}[h]
  \includegraphics[width=0.45\textwidth]{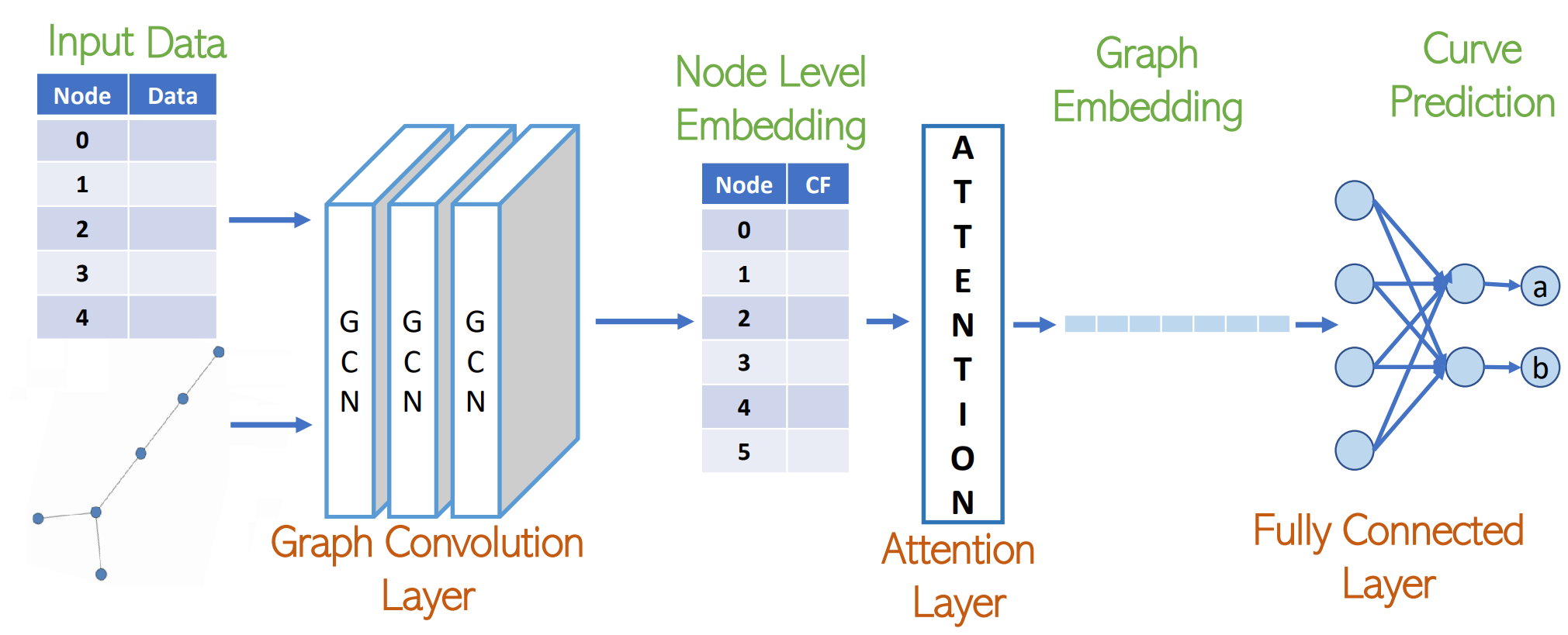}
  \vspace{-0.1cm}
  \caption{Graph Neural Network architecture}
  \vspace{-0.3cm}
  \label{fig:gnn}
\end{figure}

\textbf{Graph Neural Networks} 
We train GNN on operator level features and graph adjacency matrix to predict PCC parameters. We implement a GNN architecture, that is similar to SimGNN~\cite{bai2019simgnn}, which is computationally feasible and efficient, but also takes the node importance into consideration with an effective attention mechanism. Because of the attention mechanism, we can overweigh and focus on the most relevant part of the graph to make accurate run-time predictions. Our GNN consists
of three stages, shown in Figure~\ref{fig:gnn}. First, the input data is passed to graph convolution networks (GCN)~\cite{kipf2016semi}, a
neighbor aggregation approach, to obtain the node level embeddings. Second,
the node embeddings are fed into an attention layer, where the attention
weight represents the node's similarity to the global context. The global context is a nonlinear transformation of the weighted average of the node embeddings (whose weight is a learnable object), and the graph
embedding is the attention weighted sum of the node embeddings. And finally, the
job level convolved embeddings are then passed to the multi-layer fully
connected NN to predict the PCC parameters. 

\subsection{Loss function for NN and GNN}
\label{subsec:loss-function}

For the NN and GNN models, we construct three loss functions using the standard mean absolute error (MAE) loss. We combine several loss components corresponding to both the PCC parameters and run-time predictions to form the three loss functions: 
\textbf{LF1}: single component loss, MAE of
the curve parameters. The parameters are scaled so that neither of the
two would dominate the loss function. By scaling these parameters in the training and scaling back in the prediction, the signs of the two predicted curve parameters are guaranteed to be different, which guarantees the monotonically non-increasing trend between run-times and token amount; \textbf{LF2}: two components loss, a second penalization term of MAE (in percentage) of run-time (at the observed token count) is added to regularize the models and improve run-time estimates; and \textbf{LF3}: three components loss, a third term of mean absolute difference (in percentage) between the
NN/GNN and XGBoost run-time predictions (at the observed token count) is added, with the idea of transfer learning to leverage the learnings from XGBoost (because XGBoost makes good run-time point predictions, see Section~\ref{subsec:model-validation-full}). The weights of the components are regarded as hyper-parameters. 
\section{Evaluation}
\label{sec:evaluation}

We ran extensive experiments to evaluate the accuracy of PCC
and to compare the effectiveness of the three ML models, namely XGBoost, NN and GNN, with three different loss functions for NN and GNN. 
We consider three metrics for model prediction: (1)~whether the predicted performance curve is monotonically non-increasing (qualitative metric),
(2)~the MAE of the curve parameters (quantitative metric), as the first component is in loss function, and
(3)~the median absolute error (in percentage) of run-time prediction.

We train the models with 85k production SCOPE jobs after anonymizing identifiable information about the jobs. Both job run-time and token utilizations have right-skewed distributions. 
The job run-time ranges from $33$ seconds to $21$ hours, with the average and median being $9.5$ and $3$ minutes.
The peak number of tokens used by jobs ranges from $1$ to $6{,}287$ tokens, with mean and median of $154$ and $54$ tokens.
We test over two datasets. 
First, we gather a ground truth dataset using a small set of 200 jobs, selected based on a job selection process, 
and re-execute them at four different token values (Section~\ref{subsec:flighting-selection}). Then, we fit the PCC to obtain the curve parameters. 
We use this dataset to validate the AREPAS simulator (Section~\ref{subsec:simulator-validation}) and further evaluate model predictions (Section~\ref{subsec:model-validation-flighted}).
Second, we consider a large set of 78k  SCOPE jobs, that were submitted a day after the training jobs (Section~\ref{subsec:model-validation-full}).
For this dataset, we have the ground truth run-time at only one token value. We use synthetic data from the AREPAS simulator as run-time at all other token values, regarding them as proxy ground truth. 

\subsection{Gathering Ground Truth Data}
\label{subsec:flighting-selection}

We re-executed production jobs to generate the ground truth dataset for validating both AREPAS and the PCC models. 
However, given that production resources are scarce, we can only re-execute a small fraction of the workloads, each with alternate token counts.
Our goal is to obtain a subset of jobs that match the population
distribution, so that results on this subset could be generalized; and to obtain a subset that contains as many unique jobs as
possible. A recent work generates summarized workload set from the workload population while maximizing coverage and representation~\cite{DIAMetrics}. However, in our case,
we have to subset from a pool of jobs that meet certain constraints, e.g., within certain time frame, having tokens within a particular range, or belonging to a specific virtual cluster. 
Therefore, we design a simple but effective stratified under sampling procedure for workload subset selection from the pre-selected pool of jobs.
Our procedure includes four steps:

\begin{enumerate}[leftmargin=10pt]
\def\labelenumi{\arabic{enumi}.}
\item
  \textbf{Job Filtering}: Filtering the overall workload population based on the constraints (i.e., virtual cluster,
  token range, time frame) and forming a pre-selected pool of jobs.
\item
  \textbf{Job Clustering}: K-means clustering over the entire population to divide
  jobs into multiple clusters, and predicting the cluster for each sample in
  the pre-selected pool of jobs.
\item
  \textbf{Stratified Sampling}: 
  Random under sampling within each cluster, corresponding
  to its cluster size proportion in the population. We further set a threshold value to limit the number of times each type of job can be selected.
\item
  \textbf{Quality Evaluation}: Performing a Kolmogorov-Smirnov (KS) test~\cite{kstest} before and after the job selection. A lower KS statistic after the job selection indicates the subset distribution is closer to the population compared with the pre-selected pool of jobs.
\end{enumerate}
\begin{figure}[t]
\centering
~\includegraphics[width=0.45\textwidth]{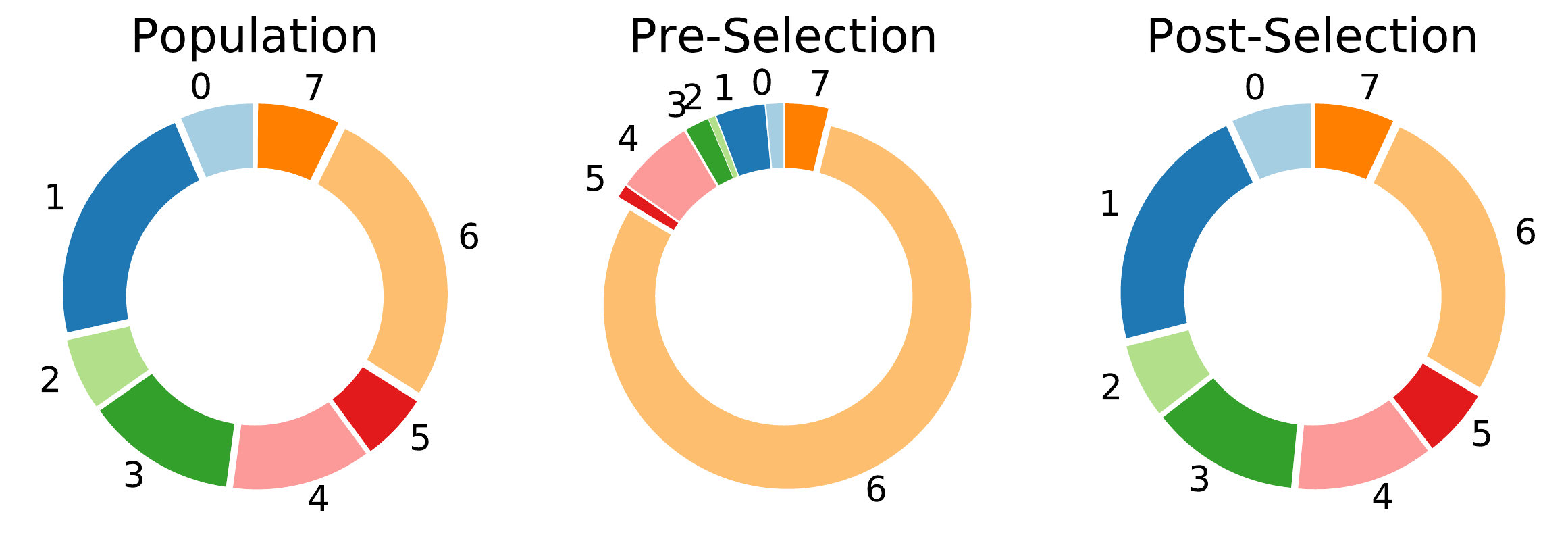}
\vspace{-0.6cm}
\caption{Clusters in the population (left), pre-selection samples (middle) and post-selection samples (right)}
\vspace{-0.4cm}
\label{fig:jselect}
\end{figure}

For validation, we selected $200$ jobs using our job selection procedure from the historical data. The population (historical jobs) is split into 8 groups and the cluster size proportion ranges from 5.9\% (group 5) to 26.7\% (group 6). In the pre-selection samples, however, the majority of (79.9\%) jobs fall in group 6 and the smallest sized group only accounts for 0.6\% (group 2). After the job subset selection, we are able to create a subset of jobs with cluster size proportion matching that of the population, as shown in Figure~\ref{fig:jselect}. We re-execute each job in the subset with four different token values, at $100\%$, $80\%$, $60\%$ and $20\%$ of the original token amount and use that as the ground truth in subsequent experiments. 

\subsection{Validating AREPAS}
\label{subsec:simulator-validation}

\begin{figure}[t]
\centering
~\includegraphics[width=0.24\textwidth]{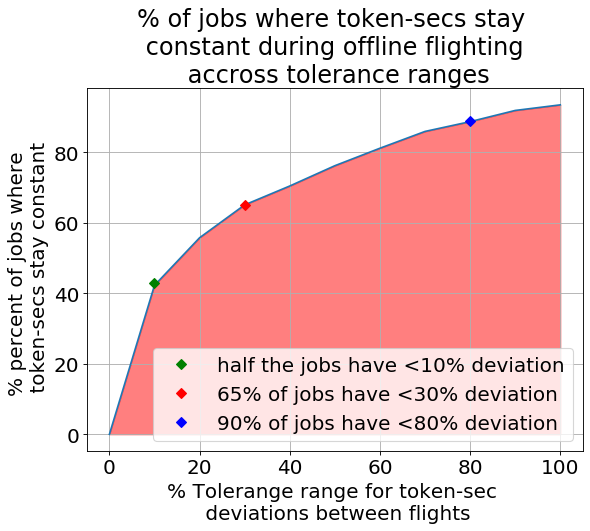}
~\includegraphics[width=0.24\textwidth]{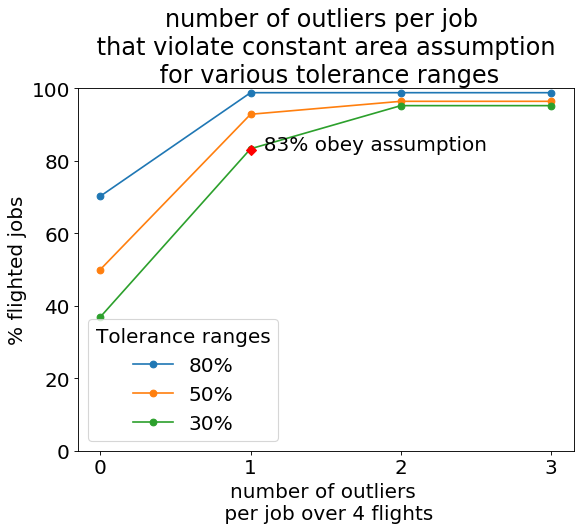}
\vspace{-0.5cm}
\caption{Validating constant resource allocation area assumption: 83\% jobs uphold the assumption within a tolerance of up to 30\% and with 1 or fewer outliers.} 
\vspace{-0.3cm}
\label{fig:validateassumption}
\end{figure}

We now validate the fundamental assumption in AREPAS, i.e., "\textit{The total token-secs or area under the resource consumption curve stays constant}". Every job is executed at 4 different token counts, and we compare the area under the resource consumption curve for each pair of executions. 4 executions per job provide ${}^{n}C_{k}=6$ execution pairs. We consider an execution pair to uphold the area conservation assumption, if the percentage difference in area is under a specified tolerance range. As we relax our tolerance by increasing the range, a greater proportion of execution pairs will be considered matches. Figure~\ref{fig:validateassumption} (\textit{left}) shows the CDF over all tolerance ranges. We observe that if we set the tolerance range to $30\%$, then $65\%$ of all execution pairs match with each other. When viewed on a granular job-by-job basis, if an execution does not match with the rest of its executions, we refer to it as an outlier. Figure~\ref{fig:validateassumption} (\textit{right}) shows the prevalence of outliers over the 4 executions of each job. 83\% of all jobs have 1 or fewer outliers over their 4 executions. Having evaluated our assumption at both the aggregate and granular level, we find AREPAS's core assumption of token-secs staying constant to be reasonable.  

\begin{figure}[t]
\centering
~\includegraphics[width=0.24\textwidth]{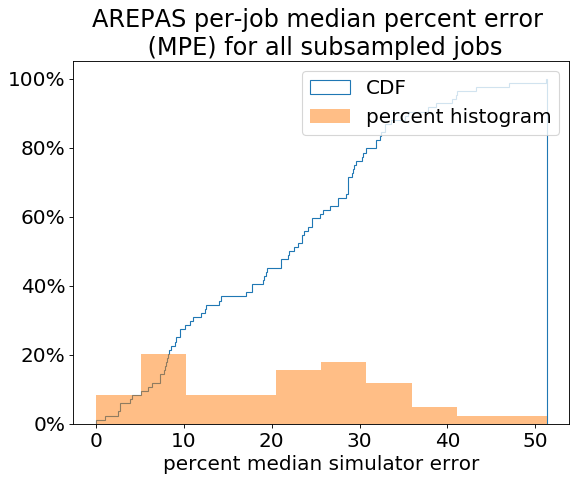}
~\includegraphics[width=0.24\textwidth]{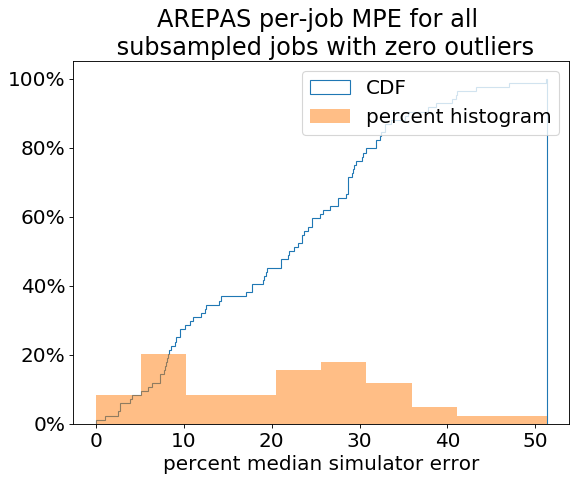}
\vspace{-0.5cm}
\caption{AREPAS accuracy against the ground truth: median error for jobs with non-anomalous behavior is 9.2\%}
\vspace{-0.3cm}
\label{fig:validateaccuracy}
\end{figure}

We now validate whether the simulations generated by AREPAS match those of the re-executed job instances. 
To focus on deterministic patterns, we discard jobs with any anomalous behavior, e.g., run-time does not decrease monotonically with more tokens. 

\begin{table}[]
\caption{AREPAS error compared to ground truth.}
\vspace{-0.2cm}
\label{tab:simulatorpe}
\small
\begin{tabular}{lccc}
\toprule
Job groups & Num. Executions & MedianAPE & MeanAPE  \\
\midrule
Non-anomalous subset & 296  & 9.19\% & 14\%   \\
Fully-matched subset & 97  & 22\% & 25\%   \\                 
\bottomrule       
\end{tabular}
\vspace{-0.5cm}
\end{table}

We also analyze a smaller (Fully-matched) subset of jobs which exhibit zero outliers on the green line as seen in Figure~\ref{fig:validateassumption} (right). These are jobs where token-secs match for all four executions. They represent 38\% of the original sample. We report Mean Average Percent Error (MeanAPE) and Median Average Percent Error (MedianAPE) for both the non-anomalous subset and the fully-matched subset in Table~\ref{tab:simulatorpe}. We observe, that AREPAS's results match the re-executed run-times quite closely, with the worst-case error being under $50\%$ and $30\%$ for the two subsets, as seen in Figure~\ref{fig:validateaccuracy}.

\subsection{Model Accuracy on Historical Dataset}
\label{subsec:model-validation-full}

\begin{table}[]
\caption{Results for 1st form of the loss function (LF1).}\vspace{-0.3cm}
\label{tab:loss1}
\small
\begin{tabular}{lccc}
\toprule
Model   & \begin{tabular}[c]{@{}l@{}}Pattern \\ (Non-Increase)\end{tabular} & \begin{tabular}[c]{@{}l@{}}MAE\\ (Curve   Params)\end{tabular} & \begin{tabular}[c]{@{}l@{}}Median AE\\ (Run-Time)\end{tabular} \\
\midrule
XGBoost SS & 41\%  & NA  & 13\%  \\
XGBoost PL & 73\%  & 0.232  & 13\%  \\
NN      & 100\%  & 0.086 & 31\% \\
GNN     & 100\%  & 0.071  & 31\%                      
        \\
\bottomrule       
\end{tabular}
\vspace{0.1cm}
\caption{Results for 2nd form of the loss function (LF2).}\vspace{-0.3cm}
\label{tab:loss2}
\begin{tabular}{lccc}
\toprule
Model   & \begin{tabular}[c]{@{}l@{}}Pattern \\ (Non-Increase)\end{tabular} & \begin{tabular}[c]{@{}l@{}}MAE\\ (Curve   Params)\end{tabular} & \begin{tabular}[c]{@{}l@{}}Median AE\\ (Run-Time)\end{tabular} \\
\midrule
XGBoost SS & 41\%  & NA  & 13\%  \\
XGBoost PL & 73\%  & 0.232  & 13\%  \\
NN      & 100\% & 0.090 & 22\% \\
GNN     & 100\% & 0.071 & 20\% \\
\bottomrule       
\end{tabular}
\vspace{0.1cm}
\caption{Results for 3rd form of the loss function (LF3).}\vspace{-0.3cm}
\label{tab:loss3}
\begin{tabular}{lccc}
\toprule
Model   & \begin{tabular}[c]{@{}l@{}}Pattern \\ (Non-Increase)\end{tabular} & \begin{tabular}[c]{@{}l@{}}MAE\\ (Curve  Params)\end{tabular} & \begin{tabular}[c]{@{}l@{}}Median AE\\ (Run-Time)\end{tabular} \\
\midrule
XGBoost SS & 41\%  & NA  & 13\%  \\
XGBoost PL & 73\%  & 0.232  & 13\%  \\
NN      & 100\% & 0.083 & 22\% \\
GNN     & 100\% & 0.077 & 21\% \\
\bottomrule       
\end{tabular}
\vspace{-0.6cm}
\end{table}

We now validate the model predictions over the historical dataset. 
Table~\ref{tab:loss1}-\ref{tab:loss3} shows the performance comparison of the three models.
For trend prediction, neither of the XGBoost models (XGBoost SS, XGBoost PL) can guarantee a
monotonically non-increasing PCC even after data augmentation.
For XGBoost SS, only 41\% of the jobs have non-increasing PCC
(within +-40\% of the reference, i.e., the observed token count), and around 60\% the predicted PCCs have at least one
region of increase. For XGBoost PL, around 73\% of the jobs have predicted PCC non-increasing (the two predicted curve parameters have different signs),
and 27\% of the jobs have the predicted PCC monotonically increasing (predicted curve parameters have the same sign),
i.e., run-time increases with token counts.
This is because point predictions for XGBoost do not necessarily decrease with the increase of tokens. For NN and GNN, a monotonically decreasing PCC
is fitted and guaranteed, because the curve parameters are scaled and treated directly as target variables. 
GNN has a smaller error (MAE) of curve parameters
(0.071--0.077), compared to NN (0.083--0.090). The MAE of curve parameters for XGBoost PL is much higher (0.232), around 3$\times$ that of NN and GNN. 

For run-time prediction at the observed token count, XGBoost has smaller error, because XGBoost models the run-time directly and aims to minimize the run-time error only. Half of the jobs have run-time prediction
error of 13\% or less. For NN and GNN, the median absolute
error (Median AE) 
is larger than for XGBoost, while the accuracy
of GNN and NN is similar.
The accuracy of run-time prediction also depends on the loss
function. The error of NN and GNN with LF1 is the
largest, because it has a single target to minimize the error of curve
parameters. For LF2, the penalization term is
added and as a result, the models aim to minimize the run-time prediction
error as well. The curve parameter loss and run-time related
loss are balanced by applying weights. We tuned the penalization
weights, so that the MAE of the curve parameters in LF2 is close to that of LF1. By
adding the penalization terms, the run-time prediction is substantially
improved for NN and GNN (31\% for LF1 vs. 
20-22\% for LF2) without sacrificing the
accuracy of curve parameters prediction. When comparing LF2 with LF3, the NN/GNN model performances do not differ
significantly. Hence, adding another penalization term to take advantage
of XGBoost's learning is redundant.

\begin{table}[]
\caption{Parameter counts, training and inference times.}
\vspace{-0.3cm}
\label{tab:nnvsgnn}
\small
\begin{tabular}{lccc}
\toprule
Model &  \begin{tabular}[c]{@{}l@{}}Number of \\ Parameters \end{tabular} & \begin{tabular}[c]{@{}l@{}}Time(s): Training\\Per epoch \end{tabular} & \begin{tabular}[c]{@{}l@{}}Time(s): Inference\\ Per 
10,000 jobs\end{tabular} \\
\midrule
NN & 2,216 & 2 & \textbf{0.09} \\
GNN & 19,210 & 913 & 78 \\
\bottomrule
\end{tabular}
\vspace{-0.6cm}
\end{table}

Among the models that we studied, XGBoost SS makes
no assumption of the shape of the PCC. At the same time, XGBoost PL makes assumption of the power-law shaped PCC as NN and GNN do. 
The XGBoost models suffer from two major limitations. Firstly, the
non-increasing pattern of the PCC is not
guaranteed. We obtain the non-increasing curve from XGBoost SS for less
than half of the jobs within a local range (+-40\%) of the reference
point. For XGBoost PL, the predicted run-time has a monotonically increasing relationship with token count for 27\% jobs.
This could cause confusion to users as an increase in job run-time with more tokens is not expected.
Secondly, the model performance outside the local range
of the reference point is not expected to perform as well. Thus, the
construction of the range as well as the prediction of the reference
point is needed.

In contrast, the NN and GNN models with LF2 are accurate for both trend and point predictions. We prefer NN/GNN to XGBoost.
They aim to learn the
shape of the PCC from AREPAS and predict run-time using the predicted power-law curve
that is monotonically decreasing for the whole range. 
Both models make assumptions of the PCC shape (specified by two parameters). 
As shown in Table~\ref{tab:nnvsgnn},
the NN is
more lightweight than the GNN model (9$\times$ fewer
parameters), is less computationally intensive (450$\times$ faster
training, 900$\times$ faster  
predictions), and does
not require the job graph data. Meanwhile, the GNN model shows
marginally better performance than NN, but requires job graph data and
more computational capacity.

\subsection{Model Accuracy on Ground Truth Data}
\label{subsec:model-validation-flighted}

We now evaluate XGBoost SS, XGBoost PL, NN and GNN with LF2 on the ground truth data and compare the predicted run-time and curve parameters with the job re-execution results. Table~\ref{tab:resultflight} shows that
the errors for XGBoost are the largest, followed by NN and GNN. 
For point prediction, the MAE of run-time for XGBoost is 4$\times$ that for the full historical jobs augmented with AREPAS (also see Table~\ref{tab:loss2}), while for NN and GNN it is 2$\times$ and $\sim$1.5$\times$ respectively. 
For trend prediction, the MAE of curve parameters for NN and GNN is 2$\times$ that of the full historical jobs, while for XGBoost PL it is smaller. NN and GNN have smaller error on both point and trend prediction, compared with XGBoost. 
Thus, although the precision of the simulator affects model accuracy, our NN and GNN models do reasonably well on the ground truth data.

\begin{table}[]
\caption{Results on ground truth data.}
\vspace{-0.3cm}
\label{tab:resultflight}
\small
\begin{tabular}{lccc}
\toprule
Model   & \begin{tabular}[c]{@{}l@{}}Pattern \\ (Non-Increase)\end{tabular} & \begin{tabular}[c]{@{}l@{}}MAE\\ (Curve  Params)\end{tabular} & \begin{tabular}[c]{@{}l@{}}Median AE\\ (Run-Time)\end{tabular} \\
\midrule
XGBoost SS & 32\%  & NA  & 53\%  \\
XGBoost PL & 93\%  & 0.202 & 52\%  \\
NN      & 100\% & 0.163 & 39\% \\
GNN     & 100\% & 0.168 & \textbf{33\%} \\
\bottomrule       
\end{tabular}
\vspace{-0.6cm}
\end{table}

\section{Related Work}
\label{sec:related-work}

Efficient resource allocation has received a lot of attention in recent years~\cite{10.1145/3267809.3267819, Alipourfard2017cherrypick, ferguson2012jockey, jyothi2016morpheus, 10.1145/2987550.2987566, Khan2016hadoopmodel, sen2020autotoken, fan2020automated}.  
For SCOPE, Jockey~\cite{ferguson2012jockey} and PerfOrator~\cite{10.1145/2987550.2987566} study efficient resource allocation using non-ML approaches. Morpheus~\cite{jyothi2016morpheus} considers past resource-usage skylines and uses linear programming for their job resource model. However, it doesn't model critical relationships such as those between the amount of data and run-time.
AutoToken~\cite{sen2020autotoken} builds specialized models for subsets of workloads by considering a rich set of features.
However, it does not predict allocations for ad-hoc (non-recurring) jobs and does not predict run-time or impact on run-time for given token allocations. 
A recent approach to adapt resources to the usage skyline progressively estimates peaks in the remaining lifetime of the job and releases any excess resources~\cite{tokenshaper}.
However, it cannot be more aggressive in reclaiming resources. Additionally, it requires deep integration and constant communication with the job scheduler during the run-time of the job and cannot provide estimates or insights into the job's execution at compilation time.

Other works have tried to capture performance models by executing sample runs for a variety of resources, and then learning generalized linear models~\cite{Venkataraman16Ernest,Zhang13}. 
However, they either incur overheads for every new job or make linear scaling assumptions that do not hold for big data workloads. 
Likewise, others have tried to build black-box performance models for tuning Spark configurations~\cite{Wang2016sparktuning} and models to predict run-time of SQL queries for different degrees of parallelism~\cite{fan2020automated,fan2020comparative}.
TASQ, in contrast, constraints the model to the expected performance characteristics to learn a PCC and can do trend predictions using that. Furthermore, it uses a novel simulator to cheaply yet accurately augment the training data.

\section{Conclusion}

This paper pushes the envelope for resource optimization in big data analytics.
In contrast to peak resource allocation in prior works, we explore the {\it optimal} resource allocation that trades small to minimal loss in performance for more aggressive allocation to improve the overall efficiency.
We presented TASQ, an end-to-end 
learning approach to capture the performance characteristics curves (PCC)
of analytical jobs 
as a function of resource allocation. We introduced the AREPAS simulator that can
be used to augment limited ground truth data with synthetically generated skylines at different resource allocations.
We model the PCC as a power-law curve with monotonicity constraints, and describe building three kinds of models with multiple loss functions.
Our results show that TASQ can easily augment large workload with low cost and error, and do both point and trend predictions within $22\%$ median error.

\bibliographystyle{habbrv}
\balance
\bibliography{refs}

\end{document}